\documentclass[twocolumn,amsmath,amssymb,floatfix]{revtex4}

\usepackage{graphicx}
\usepackage{dcolumn}
\usepackage{bm}
\usepackage{subfig}
\usepackage{caption}
\begin{document}

\title{Ion Acoustic Waves in  Ultracold Neutral Plasmas}

\author{J. Castro, P. McQuillen, and T.C. Killian}
\affiliation{%
Rice University, Department of Physics and Astronomy and Rice Quantum Institute, Houston, Texas 77005.}%
\date{\today}

\begin{abstract}

We photoionize laser-cooled atoms with a laser beam possessing spatially periodic intensity modulations to create ultracold neutral plasmas with controlled density perturbations.
Laser-induced fluorescence imaging reveals that the density perturbations oscillate in space and time, and the dispersion relation of the oscillations matches that of ion acoustic waves, which are long-wavelength, electrostatic, density waves.
\end{abstract}


\maketitle

Collective wave phenomena are central to the transport and thermodynamic properties of plasmas, and the presence of a rich spectrum of collective modes is a distinctive feature that separates this state of matter from a simple collection of charged particles \cite{sti92}. In ultracold neutral plasmas (UNPs) \cite{kil07,kpp07}, which are orders of magnitude colder than any other neutral plasma and can be used to explore the physics of strongly coupled systems \cite{ich04,mur06PRL,mth07}, little work has been done to study collective modes \cite{fzr06,kkb00,zfr08,rha03}. Here we employ a new technique for creating controlled density perturbations to excite ion acoustic waves (IAWs) in an UNP and measure their dispersion relation.
This flexible technique for sculpting the density distribution will open  new areas of plasma dynamics for experimental study,  including the effects of strong coupling on dispersion relations \cite{rka97,mur98,kaw01,oha00PRL} and non-linear phenomena \cite{nbs99,yra74,rha03,kpp07}
in the ultracold regime.

UNPs are formed by photoionizing laser-cooled atoms near the ionization threshold. They stretch the boundaries
of traditional neutral plasma physics and have extremely
clean and controllable initial conditions that
make them ideal for studying phenomena seen in more
complex systems, such as plasma expansion and equilibration in high-energy-density laser-matter interactions \cite{mur06PRL} and quark-gluon plasmas \cite{mth07}.
UNPs have shown
fascinating dynamics, such as kinetic energy oscillations
that directly reflect the strong coupling of ions \cite{csl04,mur06PRL}.
Strong coupling arises when particle interaction energies
exceed the kinetic energy \cite{ich04}. It is important in many
fields of physics spanning classical to quantum behavior \cite {miv09,don99,mur06PRL,mth07}
and gives rise to phase transitions and the establishment
of spatial correlations of particles \cite{ich04}.
These studies complement experiments probing strong
coupling in dusty plasmas \cite{miv09} and non-neutral plasmas
of pure ions or electrons \cite{don99}.

Previous experimental studies of collective modes in UNPs were limited to excitations of Langmuir (electron density) oscillations with radio frequency electric fields \cite{fzr06,kkb00} which did not determine a dispersion relation and were relatively insensitive to dynamics of the strongly coupled ions.  A high-frequency electron drift instability was observed in an UNP in the presence of crossed electric and magnetic fields \cite{zfr08}. Spherically symmetric ion density modulations were shown to excite IAWs in numerical simulations of UNPs \cite{rha03}. Here we excite IAWs through direct imprinting of ion density modulations during plasma formation and image them in situ with time resolved laser-induced fluorescence \cite{cgk08}.


Low frequency electrostatic, or longitudinal ion density waves are one of the most fundamental oscillations in a plasma along with  Langmuir oscillations \cite{tla29}.
A hydrodynamic plasma description, assuming slow enough ion motion for electrons to remain isothermal and an infinite homogeneous medium, predicts the dispersion relation for frequency $\omega$ and wavevector $k$ \cite{sti92}
\begin{equation}
\label{eq:dispersionlong}
\left(\frac{\omega}{k}\right)^2=\frac{k_BT_e/M}{1+k^2\lambda_D^2}
\end{equation}
where $M$ is the ion mass, $T_e$ is electron temperature,
and $\lambda_D\equiv\sqrt{\epsilon_0T_e/n_ee^2}$ is the Debye screening length, for electron density $n_e$  and proton charge $e$. We have neglected an ion pressure term because the ion temperature satisfies $T_i\ll T_e$ in UNPs.
In the long wavelength limit, which is the focus of this study, this mode takes the form of an IAW with $\omega=k\sqrt{k_BT_e/M}$, in which ions provide the inertia and electrons provide the restoring pressure.

IAWs are highly Landau damped unless $T_i\ll T_e$ \cite{sti92}, 
however, they have been studied in many types of high-temperature laboratory plasmas \cite{rda60,yra74,nbs99}. Closely related to this work,  acoustic 
 waves of  highly charged dust particles in dusty plasmas have been studied experimentally \cite{bmd95,pgo96} and theoretically \cite{rka97,mur98,oha00PRL,kaw01} because of the possibility of observing the effects of strong coupling on the dispersion relation, but to date these effects have been masked by damping due to collisions with background neutral gas \cite{wbh97,rka97,kaw01}.
Beyond fundamental interest, IAWs are invoked to explain wave characteristics observed in Earth's ionosphere \cite{koe02} and transport in the solar wind, corona, and chromosphere \cite{cbe07}.

\begin{figure}[ht]
\subfloat{\label{fig:beamsetup}\includegraphics[height=1.6in]{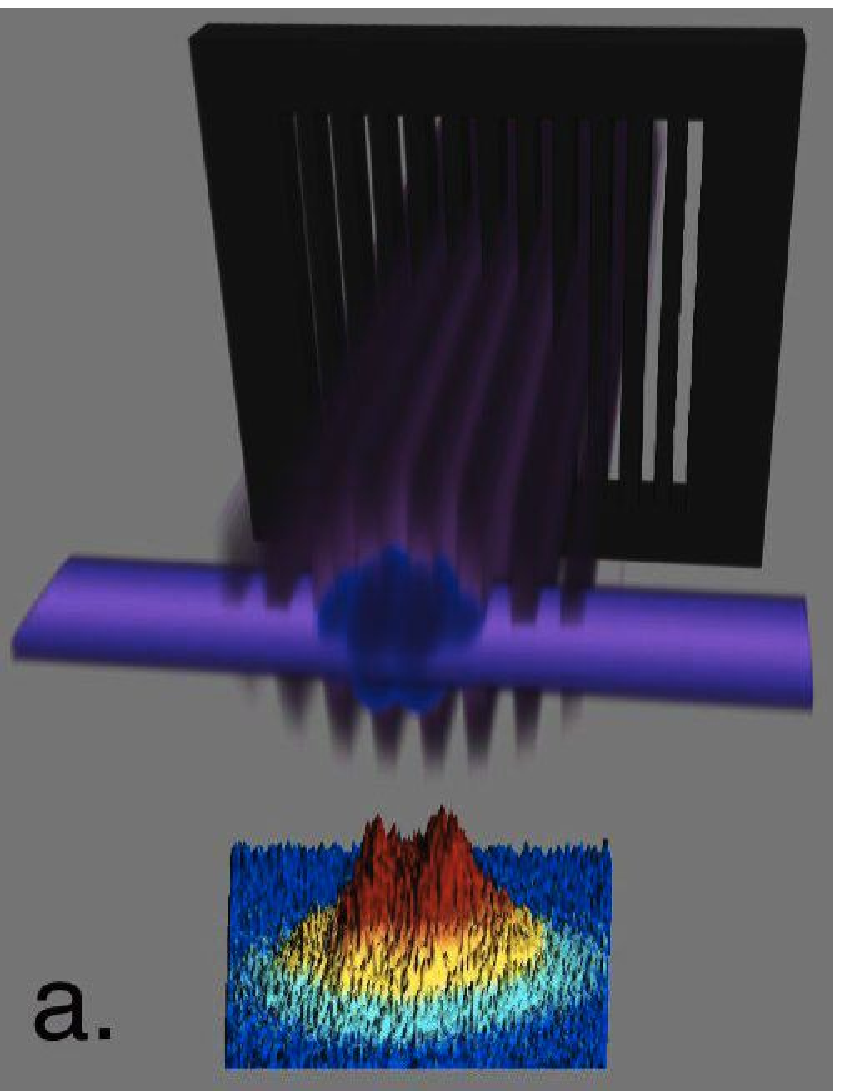}} \,
\subfloat{\label{fig:Gaussianfit}\includegraphics[height=1.6in]{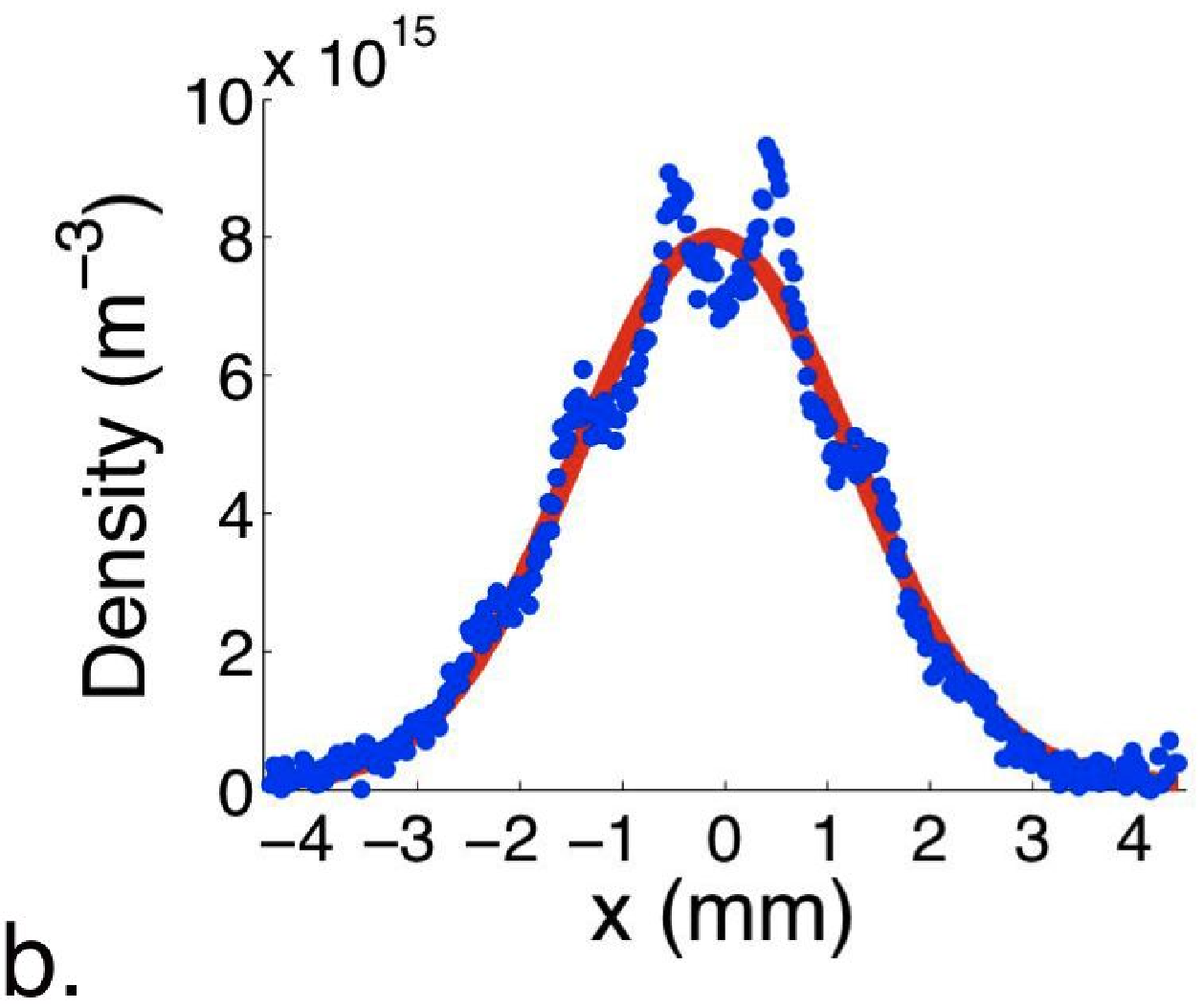}} \\
\subfloat{\label{fig:Wavepattern}\includegraphics[width=3.2in]{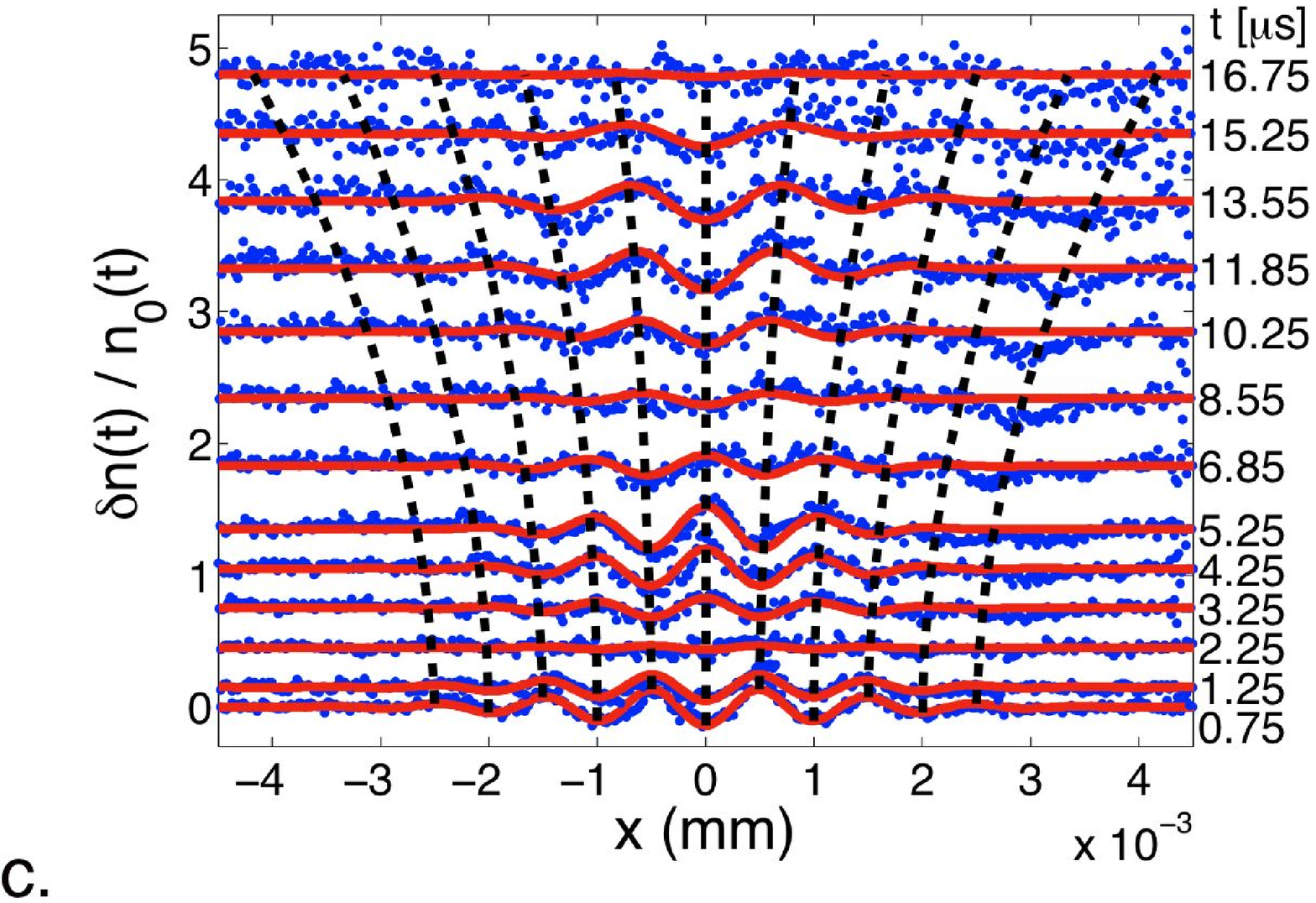}}
\caption{Excitation, imaging and modeling of density perturbations in an UNP. a, A periodic transmission mask modulates the ionization beam intensity to excite IAWs. An adjustable delay later, a sheet of laser light near resonant with the principle transition in the ions, 
illuminates a central slice of plasma, and laser-induced fluorescence along a perpendicular direction is captured by an imaging system to form a two-dimensional false color plot of the ion density distribution. b, 1D slice through the density profile for $T_e(0)=105$\,K and 1\,mm mask period, 750\,ns after ionization. Deviations from the fit Gaussian represent the IAW density modulation, $\delta n$. c, Evolution of  $\delta n$ for the same initial conditions as figure \ref{fig:ExpSetup}b. Time since ionization is indicated on the right, and $\delta n$ has been scaled by instantaneous peak density $n_0(t)$ and offset for clarity. Solid red lines are fits to equation (\ref{eq:wavemodel}), and the dotted black  lines follow the wave nodes and antinodes with time.}
\label{fig:ExpSetup}
\end{figure}

UNPs are created through photoionization of laser-cooled strontium atoms from a magneto-optical trap (MOT) \cite{kpp07}. The MOT operates on the $^{88}$Sr $ ^{1}$S$_{0}-^{1}$P$_{1}$ resonance at 461 nm \cite{nsl03}, trapping $\sim3\times10^8$ atoms at $\sim10$ mK with a spherical Gaussian density distribution. Photoionization is a two-photon process: the first from a pulse amplified laser beam operating on the Sr trapping transition, and the second from a 10 ns Nd:YAG pumped dye laser tunable around 412 nm \cite{kpp07}. This process ionizes $\sim$50\% of the atoms, and the plasma inherits its density distribution from the neutral atoms. Effects of un-ionized atoms on the plasma are not observed because of the fast time-scales of the experiment and small neutral-ion collision cross-sections.

Resulting electron temperatures are determined by the excess energy of the ionizing photons above threshold, and are adjustable from 1-1000\,K. Ion temperatures of approximately 1\,K  are set by disorder-induced heating \cite{mur01,csl04,cdd05} during the thermalization of the ions, and result in strongly-coupled ions in the liquid-like regime \cite{kpp07,ich04}. Density distributions are spherical Gaussians, $n(r)=n_0\mathrm{exp}(-r^2/2\sigma_0^2)$, with  $n_0\sim 10^{15}$\,cm$^{-3}$ and $\sigma_0\sim 1.5$\,mm, yielding average $\lambda_D$ from 3-30\,$\mu$m. In our experiments, no magnetic field is applied.

 IAWs were excited by passing the 412 nm ionizing beam, after its first pass through the plasma, through a periodic transmission mask and retroreflecting it back onto the plasma. This creates a $\sim10\%$ plasma density modulation with wavelength set by the period ($\lambda_0$) of the mask (Fig.\ \ref{fig:ExpSetup}a). The mask pattern is translated to align a density minimum to the center of the plasma.
Small higher-harmonic IAWs, arising from the square-wave nature of the mask, were observed for longer period gratings, but no effect on the fundamental wave was detected.


 For a diagnostic,
 ions are optically excited on the primary Sr$^+$ transition, $^{2}$S$_{1/2}-^{1}$P$_{1/2}$, with a tuneable, narrowband  ($\sim 5$\,MHz) laser at $\lambda$ = 422 nm, propagating approximately perpendicular to the ionizing laser \cite{kpp07,cgk08}. The 422\,nm beam is masked with a 1\,mm slit so it only
illuminates a central slice of the plasma, and resulting laser-induced fluorescence emitted close to perpendicular to the plane of the ionizing and 422\,nm beams is imaged onto an intensified CCD camera with a resolution of $13\,\mu$m.

Data from $\sim$ 50 repetitions of the experiment is summed to form a single image, $F(x,y,\nu)$, that has a frequency ($\nu$) dependence reflecting the natural linewidth and Doppler-broadening of the transition \cite{kpp07,cgk08}.  40 images are recorded
at evenly-spaced frequencies for the 422\,nm laser spanning the full spectral width of the signal. These are summed to obtain a signal proportional to the density of the plasma in the illuminated plane ($z\approx 0$),
\begin{equation}
\label{eq:Fluorimage}
 \sum_\nu F(x,y,\nu)\propto n_i(x,y,z\approx 0).
\end{equation}
Density averaging along the imaging axis ($z$) is small because fluorescence is only excited in a sheet of plasma. Absolute determination of density is obtained by calibrating fluorescence signals against absorption images of the plasma \cite{kpp07,cgk08}.

Figure \ref{fig:ExpSetup}b shows a central 1D  slice through density data along the modulation direction. 
The difference between a fit Gaussian distribution and the data yields the density perturbation $\delta{n}$ as shown in Fig. \ref{fig:ExpSetup}c. Note the oscillation of the wave in time and the changing wavelength of the excitation.

The perturbation has a negligible effect on the global dynamics of the plasma, which is dominated by expansion into the surrounding vacuum. For an initial Gaussian density distribution and conditions such as used here in which inelastic electron-ion collisions are negligible, the expansion is  self-similar and the characteristic plasma size changes according to \cite{kpp07}
\begin{eqnarray}
\label{eq:ExpSize}
\sigma(t)& =& \sigma_0\left(1+ t^2/\tau_{exp}^2\right)^{1/2},
\end{eqnarray}
where $\tau_{exp}\approx\sqrt{M\sigma_0^2/k_BT_e(0)}$ is the characteristic plasma expansion time, which ranges from 10-30 $\mu$s in this study. Associated with expansion is an adiabatic cooling of electrons according to $T_{e}(t)  =  T_{e}(0)/(1+ t^2/\tau_{exp}^2)$, which shows that expansion is driven by a transfer of  electron thermal energy to the kinetic energy of ion expansion.  In direct measurements of plasma size and spectral measurement of ion expansion velocity \cite{cgk08}, we saw no deviation in expansion dynamics due to the presence of  density modulations.

To describe wave data such as Fig.\ \ref{fig:ExpSetup}c, we scale the amplitude to the peak density at $t=0$  of the Gaussian fit (Fig.\ \ref{fig:Gaussianfit}) and assume a
 damped standing wave with a Gaussian envelope,
\begin{eqnarray}
\frac{\delta{n}}{n(0)} & = & \frac{\delta{n_0}}{n(0)}e^{-x^2/2\sigma_{env}(t)^2}\cos{[k(t)x]}\cos{[\phi(t)]\mathrm{e}^{-\Gamma t}} \nonumber \\
\label{eq:wavemodel}
 & = & A(t)e^{-x^2/2\sigma_{env}(t)^2}\cos{[k(t)x]}.
\end{eqnarray}
This expression fits the modulation at a single time, $t$,  and yields the instantaneous amplitude $A(t)$, envelope size $\sigma_{env}(t)$ and wavevector $k(t)$.  
The hydrodynamic description for an infinite homogeneous medium used to derive the IAW dispersion relation, Eq.\ \ref{eq:dispersionlong}, allows for planar standing-wave solutions, but finite size, plasma expansion, and density inhomogeneity introduce additional factors that are not small here and preclude an analytic solution, so this model is only phenomenological.

\begin{figure}[ht]
\begin{center}
\includegraphics[width=3.2in]{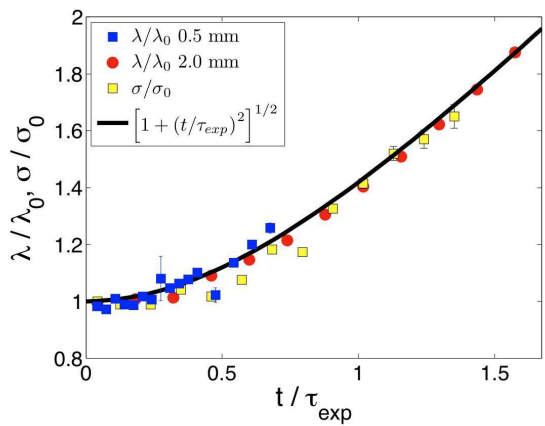}
\caption{Evolution of the IAW wavelength and plasma
size, normalized to initial values, for $T_e(0) = 70$\,K and $\sigma_0=1.45$\,mm.
Both quantities follow a universal curve indicating the
wave vector expands with the plasma. For the solid
line, $\tau_{exp}$ has been set to its theoretical value.}
\label{fig:KvectorEvol}
\end{center}
\end{figure}

To examine the effect of the expanding plasma on the wave we compare the change in the wavelength with the changing size of the plasma by plotting, in Fig.\ \ref{fig:KvectorEvol}, $\lambda(t)/ \lambda_0=k_0/ k(t)$ and $\sigma(t) /\sigma_0$, where all quantities have been normalized to the values at $t=0$. Initial wavelengths match the period of the mask used. All data follow one universal curve, $\left({1+ t^2/\tau_{exp}^2}\right)^{1/2}$, with $\tau_{exp}$ corresponding to $\sigma_0$ and $T_e(0)$ as indicated in Eq.\ \ref{eq:ExpSize}. The agreement for the wavelength indicates the wave is pinned to the expanding density distribution.

\begin{figure}[ht]
\begin{center}
\includegraphics[width=3.2in]{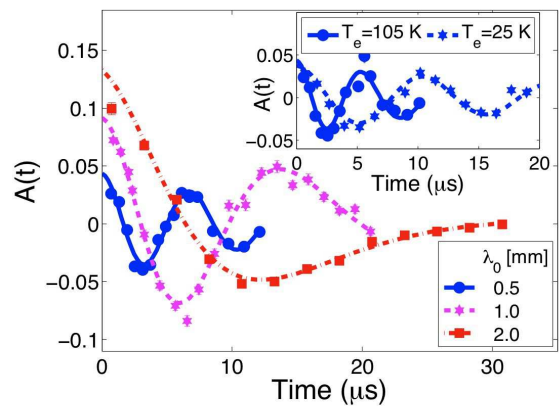}
\caption{Evolution of wave amplitude and fits to obtain $\mathbf{\omega(t)}$. The main plot shows the instantaneous amplitude for various mask periods with $T_e(0)=70$\,K.  The inset similarly compares data for different initial electron temperatures with $\lambda_0=0.50$\,mm. Lines are fits to equation (\ref{eq:ampevol}) in which $\tau_{exp}$ has been fixed to the theoretical value.}
\label{fig:AmplitudeEvolution}
\end{center}
\end{figure}

To extract the frequency of the wave, $\omega(t)$, we fit the amplitude variation in time, accounting for the  evolution of the phase, as
\begin{equation}
\label{eq:ampevol}
A(t)=A_0\mathrm{e}^{-\Gamma t}\cos\left[\phi (t)\right] =A_0\mathrm{e}^{-\Gamma t}\cos\left(\int^t_0\omega(t')dt'\right).
\end{equation}
We assume the form of the dispersion for an infinite, homogeneous medium from Eq.\ \ref{eq:dispersionlong}, the observed variation of wave-vector $k(t)$, and electron temperature evolution $T_e(t)$ predicted for a self similar expansion \cite{kpp07} to obtain
\begin{eqnarray}
\omega(t) & = & k(t)\sqrt{ k_BT_e(t)/M} = \omega_0\left(\frac{1}{1+ t^2/\tau_{exp}^2}\right).
\label{eq:omegaevol}
\end{eqnarray}
Initial amplitude $A_0$ and frequency $\omega_0$, and damping rate $\Gamma$ are allowed to vary in the fits, which match the data very well (Fig.\ \ref{fig:AmplitudeEvolution}). Decreasing frequency with time was observed in simulations of spherical IAWs \cite{rha03}.

\begin{figure}[ht]
\begin{center}
\includegraphics[width=3.2in]{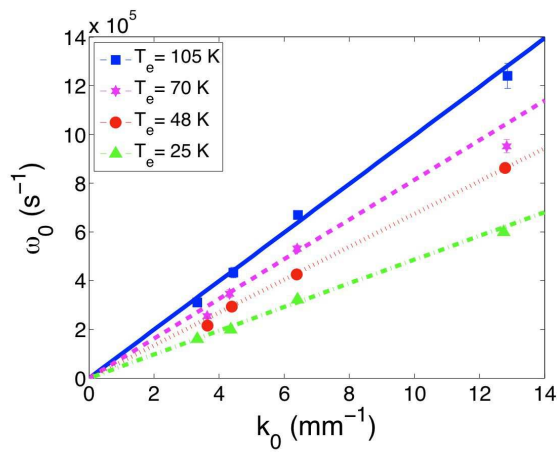}
\caption{{Dispersion relation of IAWs for different initial electron temperatures.} Lines are from the theoretical dispersion relation, Eq.\ \ref{eq:dispersionlong}, with no fit parameters.}
\label{fig:DispersionRelation}
\end{center}
\end{figure}


For a range of initial electron temperatures and mask periods, we extract $\omega_0$ and $k_0$ and calculate the dispersion of the excitations, as shown in Fig.\ \ref{fig:DispersionRelation}. The excellent agreement with theory, Eq.\ \ref{eq:dispersionlong}, confirms that these excitations are IAWs. The planar standing-wave model captures the dominant behaviour of the wave, and to a high accuracy
there is no deviation from the standard dispersion relation in spite of the plasma's finite
size, expansion, and inhomogeneous density.

Following on this initial study of ion density excitations in an UNP, there are many topics to explore. The observed waves show damping times on the order a few oscillation periods, which is faster than predicted for Landau damping \cite{sti92}. The nature of the boundary conditions and effects of density inhomogeneity and plasma expansion on IAWs should be explored, and may be important for understanding damping.
The mask period is currently limited by diffraction effects, but this can be overcome by modifying the optical configuration. For mask periods approximately five time smaller, we can probe beyond the acoustic region of the dispersion relation.
In this regime, the effects of strong coupling on dispersion are also predicted to be important, and there are many theoretical predictions that have not been tested \cite{rka97,mur98,kaw01,oha00PRL}.
The wave can be studied in velocity space with resolved fluorescence spectroscopy \cite{cgk08}, and different initial density distributions can be designed to investigate solitons \cite{nbs99}, instabilities \cite{yra74}, asymmetric excitations, and shock waves \cite{rha03,kpp07}.

This work was supported by the David and Lucille Packard Foundation and the Department of Energy and National Science Foundation (PHY-0714603).


\begin{thebibliography}{30}
\expandafter\ifx\csname natexlab\endcsname\relax\def\natexlab#1{#1}\fi
\expandafter\ifx\csname bibnamefont\endcsname\relax
  \def\bibnamefont#1{#1}\fi
\expandafter\ifx\csname bibfnamefont\endcsname\relax
  \def\bibfnamefont#1{#1}\fi
\expandafter\ifx\csname citenamefont\endcsname\relax
  \def\citenamefont#1{#1}\fi
\expandafter\ifx\csname url\endcsname\relax
  \def\url#1{\texttt{#1}}\fi
\expandafter\ifx\csname urlprefix\endcsname\relax\def\urlprefix{URL }\fi
\providecommand{\bibinfo}[2]{#2}
\providecommand{\eprint}[2][]{\url{#2}}

\bibitem[{\citenamefont{Stix}(1992)}]{sti92}
\bibinfo{author}{\bibfnamefont{T.~H.} \bibnamefont{Stix}},
  \emph{\bibinfo{title}{Waves in Plasmas}} (\bibinfo{publisher}{AIP},
  \bibinfo{address}{New York}, \bibinfo{year}{1992}), \bibinfo{edition}{2nd}
  ed.

\bibitem[{\citenamefont{Killian}(2007)}]{kil07}
\bibinfo{author}{\bibfnamefont{T.~C.} \bibnamefont{Killian}},
  \bibinfo{journal}{{Science}} \textbf{\bibinfo{volume}{316}},
  \bibinfo{pages}{705} (\bibinfo{year}{2007}).

\bibitem[{\citenamefont{Killian et~al.}(2007)\citenamefont{Killian, Pattard,
  Pohl, and Rost}}]{kpp07}
\bibinfo{author}{\bibfnamefont{T.~C.} \bibnamefont{Killian}},
  \bibinfo{author}{\bibfnamefont{T.}~\bibnamefont{Pattard}},
  \bibinfo{author}{\bibfnamefont{T.}~\bibnamefont{Pohl}}, \bibnamefont{and}
  \bibinfo{author}{\bibfnamefont{J.~M.} \bibnamefont{Rost}},
  \bibinfo{journal}{{ Phys. Rep.}} \textbf{\bibinfo{volume}{449}},
  \bibinfo{pages}{77} (\bibinfo{year}{2007}).

\bibitem[{\citenamefont{Ichimuru}(2004)}]{ich04}
\bibinfo{author}{\bibfnamefont{S.}~\bibnamefont{Ichimuru}},
  \emph{\bibinfo{title}{Statistical {P}lasma {P}hysics, Volume II: Condensed
  Plasmas}}, vol.~\bibinfo{volume}{2} of \emph{\bibinfo{series}{Frontiers in
  Physics}} (\bibinfo{publisher}{Westview Press}, \bibinfo{address}{Boulder,
  CO}, \bibinfo{year}{2004}).

\bibitem[{\citenamefont{Murillo}(2006)}]{mur06PRL}
\bibinfo{author}{\bibfnamefont{M.~S.} \bibnamefont{Murillo}},
  \bibinfo{journal}{{Phys. Rev. Lett.}} \textbf{\bibinfo{volume}{96}},
  \bibinfo{pages}{165001} (\bibinfo{year}{2006}).

\bibitem[{\citenamefont{Mrówczynski and Thoma}(2007)}]{mth07}
\bibinfo{author}{\bibfnamefont{S.}~\bibnamefont{Mrówczynski}} \bibnamefont{and}
  \bibinfo{author}{\bibfnamefont{M.~H.} \bibnamefont{Thoma}},
  \bibinfo{journal}{Annu. Rev. Nuc. and Part. Sci.}
  \textbf{\bibinfo{volume}{57}}, \bibinfo{pages}{61} (\bibinfo{year}{2007}).

\bibitem[{\citenamefont{Fletcher et~al.}(2006)\citenamefont{Fletcher, Zhang,
  and Rolston}}]{fzr06}
\bibinfo{author}{\bibfnamefont{R.~S.} \bibnamefont{Fletcher}},
  \bibinfo{author}{\bibfnamefont{X.~L.} \bibnamefont{Zhang}}, \bibnamefont{and}
  \bibinfo{author}{\bibfnamefont{S.~L.} \bibnamefont{Rolston}},
  \bibinfo{journal}{{ Phys. Rev. Lett.}} \textbf{\bibinfo{volume}{96}},
  \bibinfo{pages}{105003} (\bibinfo{year}{2006}).

\bibitem[{\citenamefont{Kulin et~al.}(2000)\citenamefont{Kulin, Killian,
  Bergeson, and Rolston}}]{kkb00}
\bibinfo{author}{\bibfnamefont{S.}~\bibnamefont{Kulin}},
  \bibinfo{author}{\bibfnamefont{T.~C.} \bibnamefont{Killian}},
  \bibinfo{author}{\bibfnamefont{S.~D.} \bibnamefont{Bergeson}},
  \bibnamefont{and} \bibinfo{author}{\bibfnamefont{S.~L.}
  \bibnamefont{Rolston}}, \bibinfo{journal}{{ Phys. Rev. Lett.}}
  \textbf{\bibinfo{volume}{85}}, \bibinfo{pages}{318} (\bibinfo{year}{2000}).

\bibitem[{\citenamefont{Zhang et~al.}(2008)\citenamefont{Zhang, Fletcher, and
  Rolston}}]{zfr08}
\bibinfo{author}{\bibfnamefont{X.~L.} \bibnamefont{Zhang}},
  \bibinfo{author}{\bibfnamefont{R.~S.} \bibnamefont{Fletcher}},
  \bibnamefont{and} \bibinfo{author}{\bibfnamefont{S.~L.}
  \bibnamefont{Rolston}}, \bibinfo{journal}{Phys. Rev. Lett.}
  \textbf{\bibinfo{volume}{101}}, \bibinfo{pages}{195002}
  (\bibinfo{year}{2008}).

\bibitem[{\citenamefont{Robicheaux and Hanson}(2003)}]{rha03}
\bibinfo{author}{\bibfnamefont{F.}~\bibnamefont{Robicheaux}} \bibnamefont{and}
  \bibinfo{author}{\bibfnamefont{J.~D.} \bibnamefont{Hanson}},
  \bibinfo{journal}{{ Phys. Plasmas}} \textbf{\bibinfo{volume}{10}},
  \bibinfo{pages}{2217} (\bibinfo{year}{2003}).

\bibitem[{\citenamefont{Rosenberg and Kalman}(1997)}]{rka97}
\bibinfo{author}{\bibfnamefont{M.}~\bibnamefont{Rosenberg}} \bibnamefont{and}
  \bibinfo{author}{\bibfnamefont{G.}~\bibnamefont{Kalman}},
  \bibinfo{journal}{Phys. Rev. E} \textbf{\bibinfo{volume}{56}},
  \bibinfo{pages}{7166} (\bibinfo{year}{1997}).

\bibitem[{\citenamefont{Murillo}(1998)}]{mur98}
\bibinfo{author}{\bibfnamefont{M.~S.} \bibnamefont{Murillo}},
  \bibinfo{journal}{Phys. Plasmas} \textbf{\bibinfo{volume}{5}},
  \bibinfo{pages}{3116} (\bibinfo{year}{1998}).

\bibitem[{\citenamefont{Kaw}(2001)}]{kaw01}
\bibinfo{author}{\bibfnamefont{P.~K.} \bibnamefont{Kaw}},
  \bibinfo{journal}{Phys. Plasmas} \textbf{\bibinfo{volume}{8}},
  \bibinfo{pages}{1870} (\bibinfo{year}{2001}).

\bibitem[{\citenamefont{Ohta and Hamaguchi}(2000)}]{oha00PRL}
\bibinfo{author}{\bibfnamefont{H.}~\bibnamefont{Ohta}} \bibnamefont{and}
  \bibinfo{author}{\bibfnamefont{S.}~\bibnamefont{Hamaguchi}},
  \bibinfo{journal}{{Phys. Rev. Lett.}} \textbf{\bibinfo{volume}{84}},
  \bibinfo{pages}{6026} (\bibinfo{year}{2000}).

\bibitem[{\citenamefont{Nakamura et~al.}(1999)\citenamefont{Nakamura, Bailung,
  and Shukla}}]{nbs99}
\bibinfo{author}{\bibfnamefont{Y.}~\bibnamefont{Nakamura}},
  \bibinfo{author}{\bibfnamefont{H.}~\bibnamefont{Bailung}}, \bibnamefont{and}
  \bibinfo{author}{\bibfnamefont{P.~K.} \bibnamefont{Shukla}},
  \bibinfo{journal}{Phys. Rev. Lett.} \textbf{\bibinfo{volume}{83}},
  \bibinfo{pages}{1602} (\bibinfo{year}{1999}).

\bibitem[{\citenamefont{Yamada and Raether}(1974)}]{yra74}
\bibinfo{author}{\bibfnamefont{M.}~\bibnamefont{Yamada}} \bibnamefont{and}
  \bibinfo{author}{\bibfnamefont{M.}~\bibnamefont{Raether}},
  \bibinfo{journal}{Phys. Rev. Lett.} \textbf{\bibinfo{volume}{32}},
  \bibinfo{pages}{99} (\bibinfo{year}{1974}).

\bibitem[{\citenamefont{Chen et~al.}(2004)\citenamefont{Chen, Simien, Laha,
  Gupta, Martinez, Mickelson, Nagel, and Killian}}]{csl04}
\bibinfo{author}{\bibfnamefont{Y.~C.} \bibnamefont{Chen}},
  \bibinfo{author}{\bibfnamefont{C.~E.} \bibnamefont{Simien}},
  \bibinfo{author}{\bibfnamefont{S.}~\bibnamefont{Laha}},
  \bibinfo{author}{\bibfnamefont{P.}~\bibnamefont{Gupta}},
  \bibinfo{author}{\bibfnamefont{Y.~N.} \bibnamefont{Martinez}},
  \bibinfo{author}{\bibfnamefont{P.~G.} \bibnamefont{Mickelson}},
  \bibinfo{author}{\bibfnamefont{S.~B.} \bibnamefont{Nagel}}, \bibnamefont{and}
  \bibinfo{author}{\bibfnamefont{T.~C.} \bibnamefont{Killian}},
  \bibinfo{journal}{{ Phys. Rev. Lett.}} \textbf{\bibinfo{volume}{93}},
  \bibinfo{pages}{265003} (\bibinfo{year}{2004}).

\bibitem[{\citenamefont{Morfill and Ivlev}(2009)}]{miv09}
\bibinfo{author}{\bibfnamefont{G.~E.} \bibnamefont{Morfill}} \bibnamefont{and}
  \bibinfo{author}{\bibfnamefont{A.~V.} \bibnamefont{Ivlev}},
  \bibinfo{journal}{Rev. Mod. Phys.} \textbf{\bibinfo{volume}{81}},
  \bibinfo{pages}{1353} (\bibinfo{year}{2009}).

\bibitem[{\citenamefont{Dubin and O'Neil}(1999)}]{don99}
\bibinfo{author}{\bibfnamefont{D.~H.~E.} \bibnamefont{Dubin}} \bibnamefont{and}
  \bibinfo{author}{\bibfnamefont{T.~M.} \bibnamefont{O'Neil}},
  \bibinfo{journal}{{ Rev. Mod. Phys.}} \textbf{\bibinfo{volume}{71}},
  \bibinfo{pages}{87} (\bibinfo{year}{1999}).

\bibitem[{\citenamefont{J.~Castro and Killian}(2008)}]{cgk08}
\bibinfo{author}{\bibnamefont{J.~Castro}}, \bibinfo{author}{\bibnamefont{H.~Gao}}
  \bibnamefont{and} \bibinfo{author}{\bibfnamefont{T.~C.}
  \bibnamefont{Killian}}, \bibinfo{journal}{Journal of Plasma Physics and
  Controlled Fusion} \textbf{\bibinfo{volume}{50}}, \bibinfo{pages}{124011}
  (\bibinfo{year}{2008}).

\bibitem[{\citenamefont{Tonks and Langmuir}(1929)}]{tla29}
\bibinfo{author}{\bibfnamefont{L.}~\bibnamefont{Tonks}} \bibnamefont{and}
  \bibinfo{author}{\bibfnamefont{I.}~\bibnamefont{Langmuir}},
  \bibinfo{journal}{{ Phys. Rev.}} \textbf{\bibinfo{volume}{33}},
  \bibinfo{pages}{195} (\bibinfo{year}{1929}).

\bibitem[{\citenamefont{Rynn and D'Angelo}(1960)}]{rda60}
\bibinfo{author}{\bibfnamefont{N.}~\bibnamefont{Rynn}} \bibnamefont{and}
  \bibinfo{author}{\bibfnamefont{N.}~\bibnamefont{D'Angelo}},
  \bibinfo{journal}{Rev. Sci. Instrum.} \textbf{\bibinfo{volume}{31}},
  \bibinfo{pages}{1326} (\bibinfo{year}{1960}).

\bibitem[{\citenamefont{Barkan et~al.}(1995)\citenamefont{Barkan, Merlino, and
  D'Angelo}}]{bmd95}
\bibinfo{author}{\bibfnamefont{A.}~\bibnamefont{Barkan}},
  \bibinfo{author}{\bibfnamefont{R.~L.} \bibnamefont{Merlino}},
  \bibnamefont{and} \bibinfo{author}{\bibfnamefont{N.}~\bibnamefont{D'Angelo}},
  \bibinfo{journal}{Phys. Plasmas} \textbf{\bibinfo{volume}{2}},
  \bibinfo{pages}{3563} (\bibinfo{year}{1995}).

\bibitem[{\citenamefont{Pieper and Goree}(1996)}]{pgo96}
\bibinfo{author}{\bibfnamefont{J.~B.} \bibnamefont{Pieper}} \bibnamefont{and}
  \bibinfo{author}{\bibfnamefont{J.}~\bibnamefont{Goree}},
  \bibinfo{journal}{Phys. Rev. Lett.} \textbf{\bibinfo{volume}{77}},
  \bibinfo{pages}{3137} (\bibinfo{year}{1996}).

\bibitem[{\citenamefont{Wang and Bhattacharjee}(1997)}]{wbh97}
\bibinfo{author}{\bibfnamefont{X.}~\bibnamefont{Wang}} \bibnamefont{and}
  \bibinfo{author}{\bibfnamefont{A.}~\bibnamefont{Bhattacharjee}},
  \bibinfo{journal}{Phys. Plasmas} \textbf{\bibinfo{volume}{4}},
  \bibinfo{pages}{3759} (\bibinfo{year}{1997}).

\bibitem[{\citenamefont{Koepke}(2002)}]{koe02}
\bibinfo{author}{\bibfnamefont{M.~E.} \bibnamefont{Koepke}},
  \bibinfo{journal}{{Phys. Plasmas}} \textbf{\bibinfo{volume}{9}},
  \bibinfo{pages}{2420} (\bibinfo{year}{2002}).

\bibitem[{\citenamefont{Cranmer et~al.}(2007)\citenamefont{Cranmer, van
  Ballegooijen, and Edgar}}]{cbe07}
\bibinfo{author}{\bibfnamefont{S.~R.} \bibnamefont{Cranmer}},
  \bibinfo{author}{\bibfnamefont{A.~A.} \bibnamefont{van Ballegooijen}},
  \bibnamefont{and} \bibinfo{author}{\bibfnamefont{R.~J.} \bibnamefont{Edgar}},
  \bibinfo{journal}{Astrophys. J.} \textbf{\bibinfo{volume}{171}},
  \bibinfo{pages}{520} (\bibinfo{year}{2007}).

\bibitem[{\citenamefont{Murillo}(2001)}]{mur01}
\bibinfo{author}{\bibfnamefont{M.~S.} \bibnamefont{Murillo}},
  \bibinfo{journal}{{Phys. Rev. Lett.}} \textbf{\bibinfo{volume}{87}},
  \bibinfo{pages}{115003} (\bibinfo{year}{2001}).

\bibitem[{\citenamefont{Cummings et~al.}(2005)\citenamefont{Cummings, Daily,
  Durfee, and Bergeson}}]{cdd05}
\bibinfo{author}{\bibfnamefont{E.~A.} \bibnamefont{Cummings}},
  \bibinfo{author}{\bibfnamefont{J.~E.} \bibnamefont{Daily}},
  \bibinfo{author}{\bibfnamefont{D.~S.} \bibnamefont{Durfee}},
  \bibnamefont{and} \bibinfo{author}{\bibfnamefont{S.~D.}
  \bibnamefont{Bergeson}}, \bibinfo{journal}{{ Phys. Rev. Lett.}}
  \textbf{\bibinfo{volume}{95}}, \bibinfo{pages}{235001}
  (\bibinfo{year}{2005}).

\bibitem[{\citenamefont{Nagel et~al.}(2003)\citenamefont{Nagel, Simien, Laha,
  Gupta, Ashoka, and Killian}}]{nsl03}
\bibinfo{author}{\bibfnamefont{S.~B.} \bibnamefont{Nagel}},
  \bibinfo{author}{\bibfnamefont{C.~E.} \bibnamefont{Simien}},
  \bibinfo{author}{\bibfnamefont{S.}~\bibnamefont{Laha}},
  \bibinfo{author}{\bibfnamefont{P.}~\bibnamefont{Gupta}},
  \bibinfo{author}{\bibfnamefont{V.~S.} \bibnamefont{Ashoka}},
  \bibnamefont{and} \bibinfo{author}{\bibfnamefont{T.~C.}
  \bibnamefont{Killian}}, \bibinfo{journal}{{ Phys. Rev. A}}
  \textbf{\bibinfo{volume}{67}}, \bibinfo{eid}{011401(R)}
  (\bibinfo{year}{2003}).

\end{thebibliography}

\end{document}